# QoS Routing using OLSR with Optimization for Flooding

[1]Suman Banik, [2]Bibhash Roy, [3]Parthi Dey, [4]Nabendu Chaki, [5]Sugata Sanyal

[1]Department of Election, Govt. of Tripura, India
email: suman.banik07@gmail.com

[2]Tripura Institute of Technology, Narsingarh, Tripura, India
email: bibhashroy10@yahoo.co.in

[3]Hindi Higher Secondary School, Agartala, Tripura, India
email: parthi.dey@gmail.com

[4Corresponding Author] University of Calcutta, Kolkata, India
email: nabendu@ieee.org

[5]Tata Institute of Fundamental Research, Mumbai, India
email: sanyal@tifr.res.in

**ABSTRACT**

Mobile Ad-hoc Network (MANET) is the self organizing collection of mobile nodes. The communication in MANET is done via a wireless media. Ad hoc wireless networks have massive commercial and military potential because of their mobility support. Due to demanding real time multimedia applications, Quality of Services (QoS) support in such infrastructure less networks have become essential. QoS routing in mobile Ad-Hoc networks is challenging due to rapid change in network topology. Consequently, the available state information for routing is inherently imprecise. QoS routing may suffer badly due to several factors including radio interference on available bandwidth, and inefficient flooding of information to the adjacent nodes. As a result the performance of the network degrades substantially. This paper aims at the solution for energy efficient QoS routing by best utilization of network resources such as energy and bandwidth. A comparative study shows that despite the overhead due to QoS management, this solution performs better than classical OLSR protocol in terms of QoS and efficient utilization of energy.

## 1. Introduction

A Mobile Ad-hoc Network (MANET) is defined by the MANET Working Group as *"an autonomous system of mobile routers (and associated hosts) connected by wireless links – the union of which forms an arbitrary graph"* [14]. It is a self-organizing dynamic multi-hop wireless networks established by a group of mobile nodes on a shared wireless channel [1]. The QoS issues such as end-to-end delay, available bandwidth, cost, loss probability, and error rate have been widely addressed in the context of internet [2, 3]. However, these have limited applications due to the bandwidth constraint and variant topology in MANET. In spite of these limitations, some QoS routing protocol have been proposed [4, 5, 7, 8]. Such protocols are on-demand in nature where the QoS requirements are known before routing. The Link State Routing is more suitable to guarantee QoS up to a large extent in MANET as the detailed information about the connectivity is available in each participating node. Thus, it increases the chances that a node will generate a route that meets the specified set of QoS routing constraints. However, in LSR, a large amount of information needs to be stored in the nodes. As a result, substantial amount of power is required for the devices. In recent years, a number of power-aware metrics have been proposed [10, 11, 12, 13]. The majority of these metrics has been applied to DSR routing protocol, so, an energetic evaluation of another protocol, i.e. the proactive protocol OLSR, arrived to the RFC status. In particular, the energy behavior of OLSR protocol has been evaluated and a novel energy aware Multi Point Relay selection mechanism has been proposed with QoS support.

## 2. Optimal Link State Routing Protocol

Optimized Link State Routing (OLSR) is a table driven proactive routing protocol for MANET. It is an optimization of link-state routing. In a classic link-state algorithm, link-state information is flooded throughout the network. OLSR uses this approach as well, but since the protocol runs in wireless multi-hop scenarios the message flooding in OLSR is optimized to preserve bandwidth. The optimization is based on a technique called Multipoint Relaying. The nodes are free to move randomly and organize themselves arbitrarily and treating each mobile host as a router. In this all the nodes contain pre-computed routes information about all the other nodes in network. This information is exchanged by protocol messages after periodic time. OLSR performs hop-by-hop routing, where each node uses its most recent topology information for routing. Each node selects a set of its neighbor nodes as MPRs (Multi Point Relays). Only those nodes selected as MPRs, are responsible for forwarding the Control Traffic. MPRs are selected such that 2-hop neighbors can be reached through at least one MPR node and OLSR provide shortest path routes to all destinations by providing link-state information for their MPR selectors. Nodes which have been selected as MPRs by some neighbor nodes announce this information periodically in their Control Messages. MPRs are used to form the route from starting node to destination node in MANET. All this information is announces to neighboring MPRs through Control

Messages. The purpose of selecting MPR is to reduce flooding overhead and provide optimal flooding distance. Figure-1 shows nodes and selection of MPRs for flooding control messages.

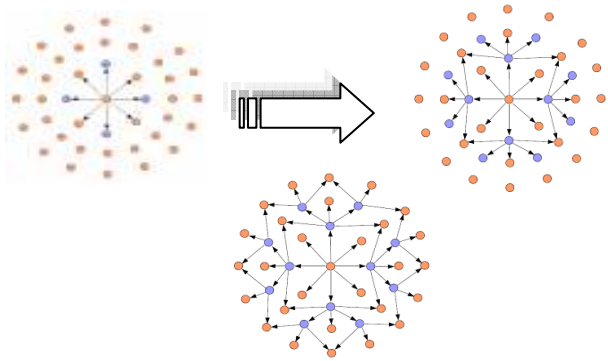

Figure-1: Flooding Optimization using OLSR

The key messages in OLSR are Hello and TC messages. Hello messages are periodically exchanged to inform nodes about their neighbors and their neighbors' neighbors and are 1-hop broadcast messages. The 2-hop neighborhood information is then used locally by each node to determine MPRs. In contrast, TC messages are flooded through the network to inform all nodes about the (partial) network topology. At a minimum, TC messages contain information about MPRs and their MPR selectors. There are few parameters in OLSR which can control the efficiency of OLSR. The Hello-interval parameter represents the frequency of generating a Hello message. Increasing the frequency of generating Hello messages leads to more frequent updates about the neighborhood and hence a more accurate view of the network and result in overhead. The TC-interval parameters represent the frequency of generating a TC message and are used for topology discovery. If frequency of TC messages is increased then nodes are having more resent information about topology, as nodes leaves and enter in the network very frequently. The MPR-coverage parameter allows a node to select redundant MPRs. The number of MPRs should be minimum as it introduce overhead in the network. But more the MPRs more is the reach ability. The TC-redundancy parameter specifies, for the local node, the amount of information that may be included in the TC message. The TC-redundancy parameter affects the overhead through affecting the amount of links being advertised as well as the amount of nodes advertising links. Through the exchange of OLSR control messages, each node accumulates information about the network. This information is stored according to the OLSR specifications. Timestamp with each data point and modify the control messages and local repositories accordingly. For better efficiency of OLSR state information such as residual energy level of each node, bandwidth, queue length etc should be available while making routing decisions. Incorrect information may lead to degradation in efficiency of OLSR. As state information in OLSR is collected by Periodic Exchange of above mentioned messages, this information may not be up to date as topology changes very fast. Residual energy level of the nodes changes rapidly and the node with less energy level must not be selected in route. The main focus here is the effect of residual energy levels on protocol efficiency.

Main thing is how nodes can collect accurate energy level information about other nodes by OLSR control messages. Traffic load can be one factor that can affect the inaccuracy of energy level information.

## 3. Integrating QoS in OLSR and Energy Constraint

QoS is a term widely used in the last recent years in the area of wire-based networks controlled by the centralized administration where fixed infrastructure is present. However it is a challenge to route QoS in wireless environment due to node's dynamic nature and mobility. The service providers implement QoS protocols keeping in mind some specific scenarios and taking into consideration different link parameters (delay, bandwidth, loss probability and error rate), network topologies and variables. In order to obtain QoS, the main emphasis must be on obtaining best bandwidth and minimum delay path. In [1], delay and hop distance are used to measure the QoS. In this proposed work, bandwidth has been considered as it is more extensively coupled with QoS routing. In [4], the focus is on reducing the link advertisement. However, the network finds it difficult to distinguish between different types of control messages. The bandwidth metric is used to specify the amount of bandwidth that will be available along the path from the initiator to the destination. In [8], the authors proposed a MPR node selection criteria based on best bandwidth path and considered the best path. Although it appears to be the optimal one, but there are factors like intrusion and radio interference in the network where in spite of having higher available bandwidth, the unexpected delay is occurred. In another work [9], the authors focused on admission control mechanism such that the bandwidth calculation is done during the routing table calculation. In this proposal, the unused bandwidth is calculated taking on account the bandwidth consumed over a link by other nodes.

## 4. OLSR Efficiency and Energy Level Accuracy

The OLSR protocol can be tuned and it can be seen that the performance changes in OLSR and how performance depends on residual energy of nodes. Some of the factors on which OLSR efficiency vary are discussed in this section. There are various MPR selection techniques and path determination algorithms available. In Modified Routing original MPR selection criteria is combined with new path determination algorithm. And in other variation Modified MPR/Routing new MPR selection and the new path determination algorithm are combined.
These variations affect performance of OLSR to a great extent. Also the protocol can be varied on the basis of "How old the information about Residual energy" is. The residual energy at that time when MPR was selected is Ideal version. In realistic version, data about residual energy collected by protocol message exchange. Also change in topology impact number of packets delivered and accuracy of the residual energy level. Packet latent also effect accuracy of data collected.

In figure-2 given below, the performance of ideal and actual version of OLSR under different traffic rate is compared. The performance of network in terms of packet delivered with respect to variation in packet interval time is compared. As Packet interval time decreases (X-Axis), more number of packets are delivered and more resent information about residual energy is collected by nodes in MANET. So inaccuracy is less and system performance increases. This is true in both ideal and realistic approach, as packet interval time decreases performances increases. But when the ideal with realistic is compared, Ideal outperforms realistic for every piece of data. It means it is sufficient to collect residual energy information at the time when MPR was selected.

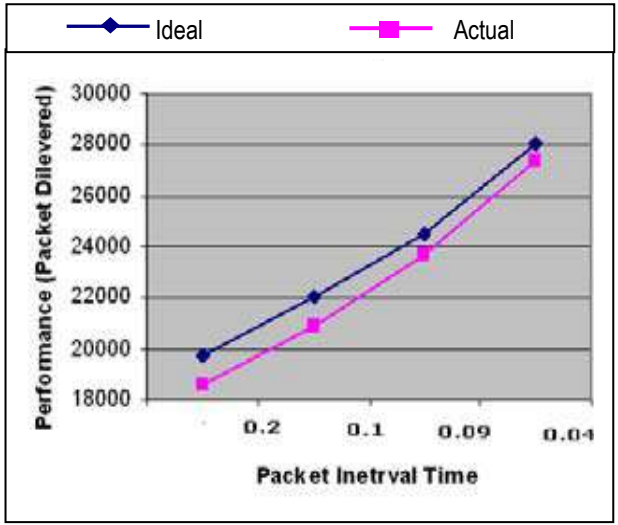

Figure-2: Ideal vs. Actual Performance

In Energy efficient variation of OLSR the MPRs are selected on the basis of residual energy levels of nodes. Path determination algorithm is modified, selecting paths based on the residual energy level of intermediate nodes. Nodes with low residual energy are avoided. The route & MPR selection is such that to maximize bottleneck residual energy level. That will increase the efficiency of network. If wrong or old information is collected by nodes then efficiency is degraded as route may vanish. But the main issue is how to collect the correct residual energy information.

One solution is use of EOLSR that select route and MPRs on basis of residual energy of nodes and number of neighbors. Ideal approach is sending more packets than realistic approach in above figure. As the traffic rate increases from low to high the Ideal approach send more and more packets. Omniscient knowledge of a node's energy level delivers more packets than the realistic version. As Choosing very small values for Hello and TC intervals will significantly increase the protocol overheads. So realistic in approach with decrease in packet interval time more and more TC and Hello messages are send in the network which increase in network overhead. That is the reason Realistic approach is little less efficient then ideal as shown in figure-2. These results are a direct consequence of the increased level of congestion in the network which results in high message loss and delay and hence less accurate state information. In figure-3 the OLSR and EOLSR is compared and it is clearly seen how energy varies with network life. With time passes by energy of nodes decay very fast. In OLSR MPRs are not frequently changed & efficiency degrade. However, in EOLSR MPRs selection depends on residual energy level of nodes. Thus EOLSR performs better then OLSR.

This study so far shows that nodes have inaccurate information about the actual residual energy levels when making routing decisions. Modifying the OLSR protocol parameters (such as increasing the Hello or TC message rates) has very limited impact on this inaccuracy.

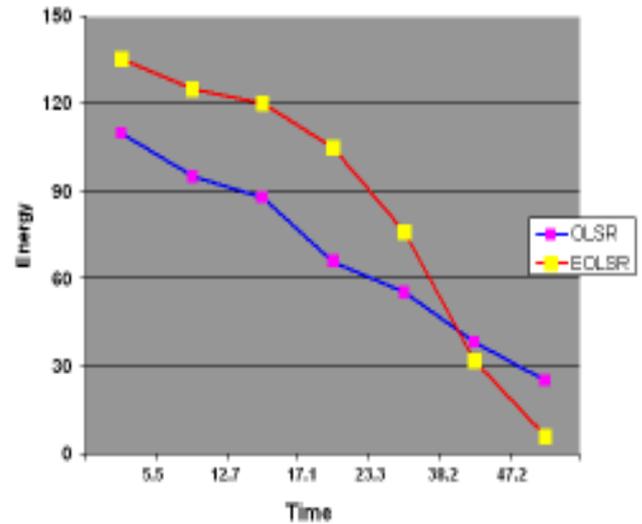

Figure-3: OLSR and EOLSR Residual Energy Levels

This means by increasing the frequency of TC and Hello messages improve residual energy information of neighboring nodes a little but increase the traffic overhead. So some other method is required to improve the accuracy of energy state information. In the next section it is suggested predictive technique to increase energy level accuracy above and beyond modifying the protocol parameters.

## 5. Reducing Inaccuracies of Residual Energy

It is clearly seen that increase in frequency of packet does not improve inaccurate energy information. So some other technique is required to compute residual energy information of nodes. In this section it is suggested that Prediction mechanism to compute residual energy information that is more accurate then previous method. Our idea is therefore to have every node locally adjust nodes' old energy levels based on their past energy consumption rate. In this mechanism each node locally extrapolates the expected energy level based on old (reported) energy levels and the energy consumption rate for that node based on the most recent two reported values. A drawback of the Prediction algorithm is the need to wait for two different perceived value readings, so a consumption rate can be calculated and used to adjust the perceived values. For predicting at least two previous values are required. If a new MPR is selected then it is not possible to predict residual energy as no previous data is available. Under high traffic loads, adjustments happen less rarely. Protocol control messages are lost / delayed, and as a result nodes will not "hear" other nodes. After a node is deemed unreachable, startup phase is again recalled, where at least two successive reports are required to be able to calculate a consumption rate.

In order to overcome the drawbacks of prediction technique smart prediction technique is used in which adjustments take place almost all the time. The number of time adjustment take place depends on packet interval times. In the Smart Prediction algorithm, for every pair of nodes (p,q), if q's consumption rate is not yet known, p adjusts the perceived value of q's residual energy level based on the average of all known consumption rates for other nodes. If p does not know a single consumption rate for other nodes, it adjusts q's perceived energy level based on its (p's) consumption rate. Using all known nodes' consumption rates eliminates the domination of outliers and ensures closeness to the actual consumption rate, assuming that nodes are somewhat homogeneous in the energy characteristics of their wireless cards. The Prediction algorithms improve the overall inaccuracy level under different traffic rates.

The improvement under higher traffic rates is not as high as it is under lower traffic rates. For an adjustment to take place, a node must have received two different reported values. But under high traffic rates, due to message loss and delays, the percentage of times adjustments take place decreases. Since the Smart Prediction algorithm addresses the problem of not being able to adjust the perceived energy level value all the time, it achieves much better performance in terms of overall inaccuracy level, especially under higher traffic rates. Both the Prediction and the Smart Prediction algorithms outperform the default OLSR protocol.

## 6. Conclusion

In MANET, state information such as residual energy level plays a major role in route selection. If latest information is not collected by nodes, then performance of the network would suffer. The effect of time at which state information was collected in ideal and in realistic approach has been evaluated. It may be inferred that even if ideal approach is better than the realistic one; the increase in frequency of packets improve the performance very little. Besides, it results in increasing traffic overhead. As a solution, prediction mechanism and smart prediction mechanism are used. This performs better than EOLSR protocol and reduce traffic load. Of course, 100% accurate state information can not be calculated due to continually changing topology. However, accuracy can be increased by using some other technique also that may be even better than prediction mechanism. In future some other methods may be suggested to compute more accurate state information.